\documentclass[twocolumn]{jpsj3}
\usepackage{amsmath,amssymb}
\usepackage{here}
\def\v#1{\mib #1}
\newcommand{\bra}[1]{\left\langle {#1} \right\vert}
\newcommand{\ket}[1]{\left\vert {#1} \right\rangle}
\def\sed{S_{\rm e}}
\def\JA{J_{\rm A}}
\def\JF{J_{\rm F}}
\def\Jl{J_{\rm L}}
\def\DCFFA{{$\rm FA\_DC$}}
\def\FFANNN{{$\rm FA\_NNN$}}
\def\FFAANNN{{$\rm FA\_ANNN$}}

\title
{
Topological Phases of Spin-1/2  Ferromagnetic--Antiferromagnetic Alternating Heisenberg Chains with Alternating Next-Nearest-Neighbour Interaction 
}

\author
{
Kazuo { Hida}\thanks{E-mail address: hida@mail.saitama-u.ac.jp}}

\inst
{Division of Material Science, Graduate School of Science and Engineering, \\ Saitama University,  Saitama 338-8570, Japan\\ 
}

\recdate
{
\today
}

\abst
{
A series of symmetry-protected topological (SPT) and trivial spin-gap phases in the  spin-1/2 ferromagnetic--antiferromagnetic alternating Heisenberg chain with alternating  next-nearest-neighbour interaction are investigated using two kinds of entanglement spectra defined by different divisions of the whole chain. In case one of the next-nearest-neighbor interactions vanishes, the model reduces to the $\Delta$-chain in which a series of spin-gap phases are found, as shown in  J. Phys. Soc. Jpn. {\bf 77}, 044707 (2008). From the degeneracy of the entanglement spectra, these phases are identified as the SPT and trivial phases. It is found that the ground-state phase boundaries are insensitive to the strength of the alternation in the next-nearest-neighbor interaction. These results are consistent with the analysis based on the nonlinear $\sigma$ model and exact solution on the ferromagnetic-nonmagnetic phase boundary.}

\begin{document}
\sloppy
\maketitle 
\section{Introduction}
The spin-gap phases in the ground states of quantum spin chains have been extensively studied in  condensed matter physics. Among them, the Haldane phase in spin-1 Heisenberg antiferromagnetic chains\cite{hal1,hal2}  
 has been  attracting renewed interest as one of the symmetry-protected topological (SPT)  phases of matter.\cite{Pollmann2010,Pollmann2012,Zang2010,Hirano2008,Chen2011} The symmetries protecting this phase are  space inversion, time reversal, and $\pi$-rotations around two axes.\cite{Pollmann2010,Pollmann2012} In general, the spin-gap phases of the quantum spin chains are classified into SPT and trivial phases in the presence of appropriate symmetry. 
It has also been proposed that these two phases can be distinguished by the even-odd parity of the degeneracy of their entanglement spectra (ES).\cite{Pollmann2010,Pollmann2012} 

A typical example of the transition between the SPT and trivial phases is the Haldane-dimer transition in the spin-1 Heisenberg chain with alternating exchange coupling $J(1\pm\delta)$.\cite{affleck-haldane,kato-tanaka,kitazawa-nomura} In this case, the transition can be intuitively understood as a rearrangement of the valence bonds. In the Haldane phase, the spin-1 sites are connected by a single valence bond, while they are connected by 0 or 2 valence bonds alternatingly in the dimer phase as  easily understood in the limit of $\delta=\pm 1$. Hence, two Gaussian transitions take place in the interval $-1 \leq \delta \leq 1$. The same feature is generalized to a spin-$S$ bond-alternating chain in which the transition takes place $2S$ times in the same interval.\cite{affleck-haldane} In the simplest case of spin-$1/2$, only one transition takes place at $\delta=0$.

In contrast to the unfrustrated case described above, it has recently been found that a ground-state phase diagram has a much complicated structure in the frustrated case even for spin-1/2. Namely, in the spin-1/2 frustrated ferromagnetic--antiferromagnetic alternating chain with next-nearest-neighbour interaction (\FFANNN),  successive phase transitions are found in which a series of trivial and SPT spin-gap phases alternate with the strength of frustration.\cite{hts,kh2016,furuya} We  speculate that the presence of these successive transitions is one of the generic features of the one-dimensional frustrated quantum ferromagnets resulting from the interplay of quantum fluctuation, frustration, and ferromagnetic correlation. 
To confirm this speculation, however, it is desirable to investigate the topological properties of similar series of phase transitions in other models. Actually, in the $\Delta$-chain with a ferromagnetic main chain and alternating ferromagnetic and antiferromagnetic interactions between the main chain and apical spins (\DCFFA), a similar series of spin-gap phases have been found.\cite{khdlt} However, the topological nature of the latter phases in the \DCFFA\ has not yet been investigated. In the present work, we regard the  \DCFFA\ as a special case of the ferromagnetic--antiferromagnetic alternating chain with alternating next-nearest-neighbour interaction (\FFAANNN) in which one of the  next-nearest-neighbour (NNN) couplings vanishes. In this parametrization,  the phase diagrams of both models turn out to be almost identical. We investigate the ES of the \DCFFA \ 
and find that it is also almost identical to that of  the \FFANNN. 
 We further investigate the ES of  the \FFAANNN \   that interpolates these two models to show that the ES is insensitive to  strength of the alternation in the NNN interaction. Our results are reproduced by the mapping onto the nonlinear $\sigma$-model\cite{furuya} and are consistent with the exact solution on the ferromagnetic--nonmagnetic phase boundary.\cite{dmitriev,dmitriev2,Suzuki2008}

This paper is organized as follows. In the next section, the model Hamiltonian is presented. In Sect. 3, the ground-state phase diagrams obtained in Refs. \citen{hts} and \citen{khdlt} are reviewed. 
In Sect. 4, the numerical results for ES are presented.  The last  section is devoted to summary and discussion.

\begin{figure}[h!]
\centerline{\includegraphics[width=7cm]{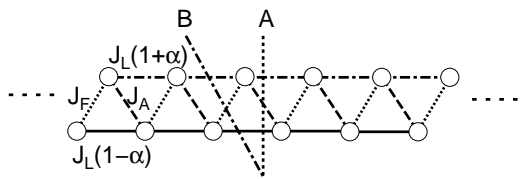}}
\caption{Lattice structure of the present system. In the calculation of the ES, the whole system is divided into left and right subsystems, as indicated by the dotted line (division A) and dash-dotted line (division B).}\label{fig:cut}
\end{figure}

\section{Model Hamiltonian}
We consider the $S=1/2$ \FFAANNN \  
described by the Hamiltonian
\begin{align}
{\cal H} &=\sum_{l=1}^{L} \JF\v{S}_{2l-1}\v{S}_{2l}+\sum_{l=1}^{L-1} \JA\v{S}_{2l}\v{S}_{2l+1}\nonumber\\
&+\sum_{l=1}^{L-1}\Jl(1+\alpha)\v{S}_{2i-1}\v{S}_{2i+1}\nonumber\\
&+\sum_{l=1}^{L-1}\Jl(1-\alpha)\v{S}_{2i}\v{S}_{2i+2}, \label{hama}
\end{align}
where $\v{S}_{i}$  is the spin-1/2 operator on the $i$-th site. In this work, we focus on the case $\JF , \Jl < 0$,  and $\JA > 0$.  The lattice structure is shown in Fig. \ref{fig:cut}.  The case $\alpha=1$ corresponds to the \DCFFA \ 
 investigated in Ref. \citen{khdlt}. The case of $\alpha=0$ has been discussed in Refs. \citen{hts} and \citen{kh2016}. For $\alpha \neq 0$, this model is no more invariant under space inversion. Nevertheless, it still has the time-reversal symmetry and $\pi$-rotations around two axes that protect the distinction between the SPT and trivial phases. Hence, the same classification of spin-gap phases as the case of $\alpha=0$ should apply. 

\section{Ground-State Phase Diagram}
\begin{figure}
\centerline{\includegraphics[width=7cm]{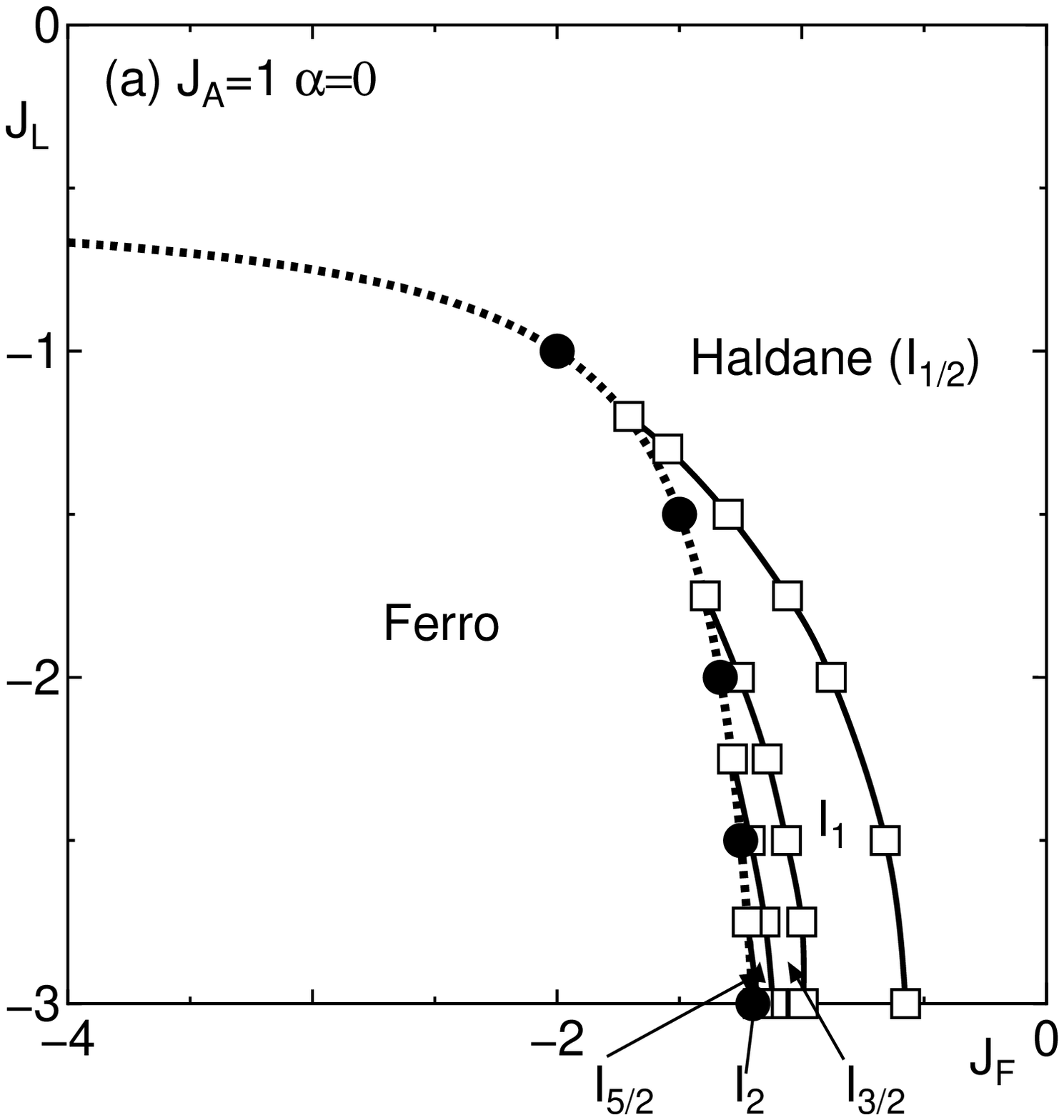}}
\centerline{\includegraphics[width=7cm]{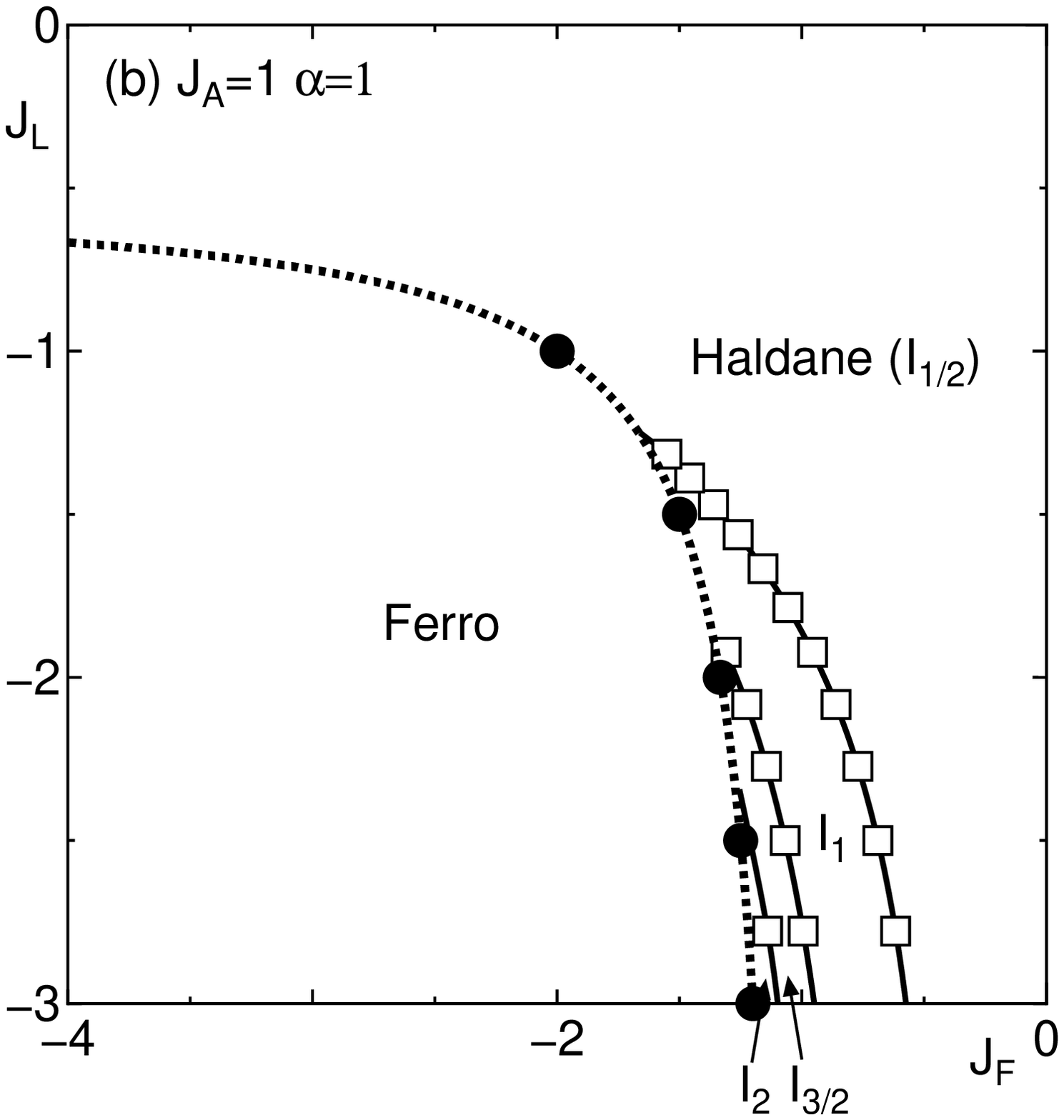}}
\caption{(a) Ground-state phase diagram for $\alpha=0$ taken from Fig. 2 of Ref. \citen{hts}.  Open squares  are the data determined  by the DMRG method.  (b) Ground-state phase diagram for $\alpha=1$ obtained by converting the data of Fig. 8 of Ref. \citen{khdlt} to present parameterization. Open squares  are the data determined  by numerical diagonalization.  In both figures, the ferromagnetic phase (F), the Haldane phase (H), and the intermediate spin-gap phases (I$_{\sed}$) are present. The dotted line is the stability limit of the ferromagnetic phase. The $I_{5/2}$-phase is not shown in (b) due to the limitation of original data.  The filled circles correspond to the ``special points'' defined by Dmitriev {\it et al.}\cite{dmitriev,dmitriev2} where exact solutions with edge spins are available.\cite{hts} The solid curves are  guides for the eye.
}
\label{fig:phase}
\end{figure}

Figure \ref{fig:phase}(a) shows the ground-state phase diagram for $\alpha=0$ taken from Ref. \citen{hts} obtained by DMRG calculation. Between the Haldane (H) phase and the ferromagnetic (F) phase, a series of spin gap phases called I$_{\sed}$ phases are present. In Ref. \citen{hts}, the I$_{\sed}$ phase is defined as the phase that has the edge spins with the magnitude $\sed$ at the two ends of the open finite chain expressed by the Hamiltonian (\ref{hama}). 

Figure \ref{fig:phase}(b) shows the ground-state phase diagram for $\alpha=1$ ($\Delta$-chain) based on the exact diagonalization data obtained in Ref. \citen{khdlt} converted into the present parametrization. Although the edge spins in the open chains were not investigated in Ref. \citen{khdlt}, the phase boundaries between the spin-gap phases are quite similar to those for $\alpha=0$. Hence, we identify the corresponding phases as the I$_{\sed}$ phases even for the case of $\alpha=1$.  
Actually, the stability limit of the ferromagnetic phase is given by
\begin{align}
   \JF&=\JF^{\rm s}\equiv-\frac{2\JA\Jl}{2\Jl+\JA},
\label{insta} 
\end{align}
irrespective of $\alpha$. On this line, the exact solution that does not depend on $\alpha$ is available.\cite{Suzuki2008} Hence, the special point solutions with edge spins\cite{hts} at $\Jl/\JA=-m/2$, where $m$ is a positive integer,\cite{dmitriev,dmitriev2} also apply irrespective of $\alpha$. This also supports the speculation that the same series of SPT and trivial phases appear within the nonmagnetic phase. In what follows, we confirm this  speculation using the ES including the case of $\alpha =0.5$.

 Recently,  the presence of these successive transitions has also been confirmed by mapping onto the nonlinear $\sigma$-model\cite{furuya} for $\alpha=0$ in the limit of large negative $\Jl$ with the topological angle $\theta$ given by
\begin{align}
\theta=-\frac{4\pi S \JF}{\JF + \JA}.
\label{eq:theta}
\end{align}
The phase boundaries are determined from the condition $\theta=(2n-1)\pi$, where $n$ is an integer. Their method can easily be extended to the case of $\alpha \neq 0$, and it turns out that the relation (\ref{eq:theta}) holds irrespective of $\alpha$. This supports our numerical conclusion that the phase boundaries are insensitive to $\alpha$.

\begin{figure}[h!] 
\centerline{\includegraphics[width=7cm]{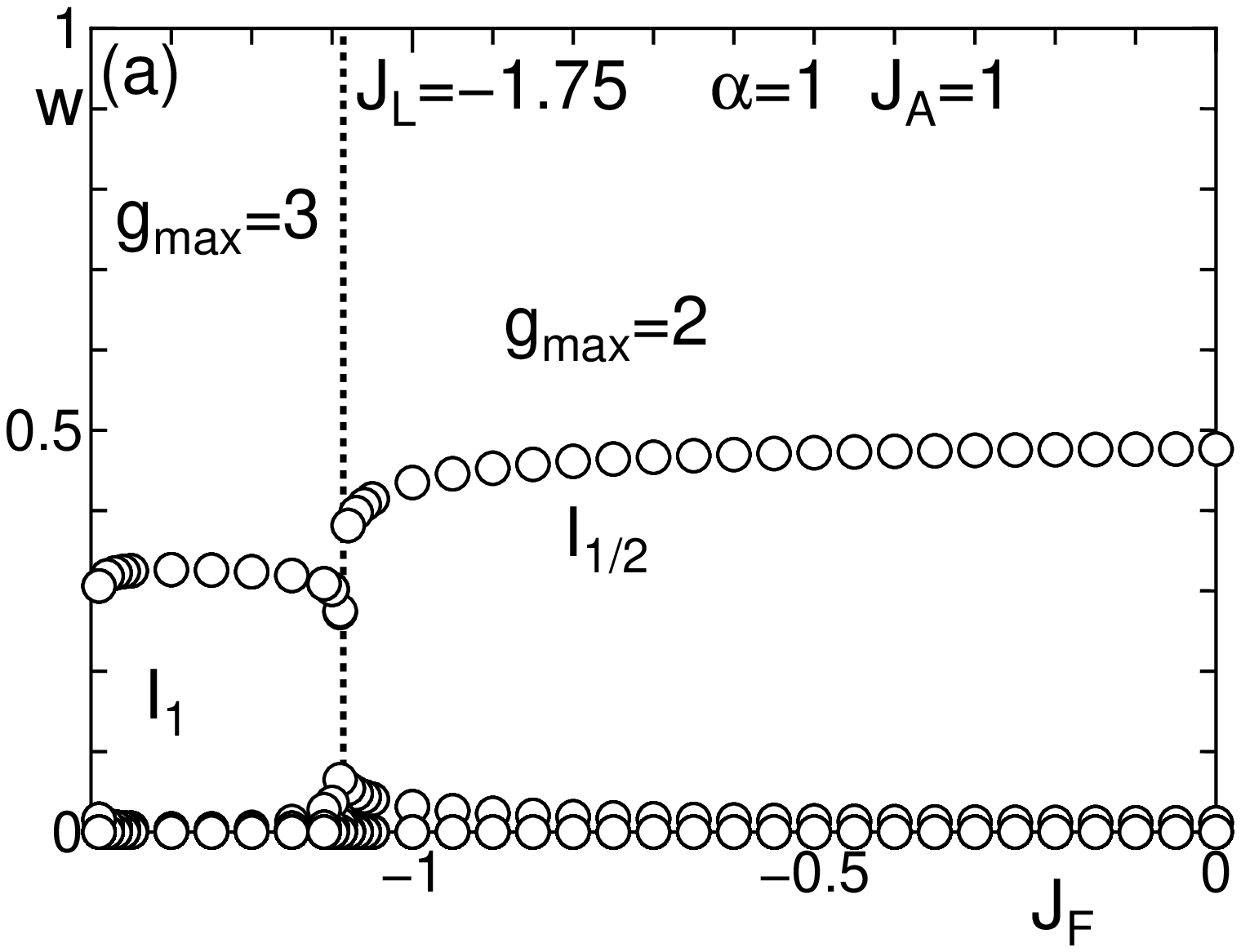}}
\centerline{\includegraphics[width=7cm]{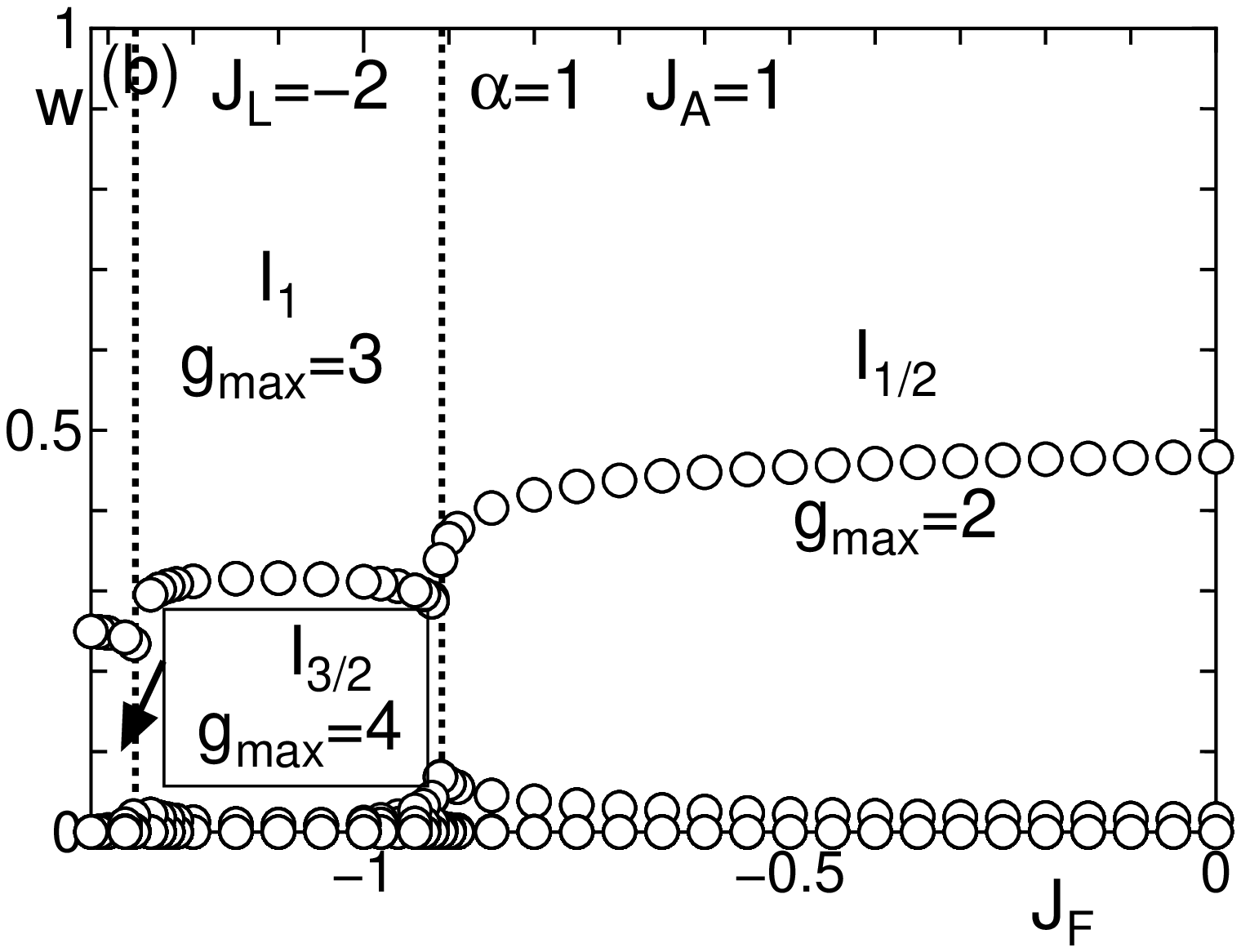}}
\centerline{\includegraphics[width=7cm]{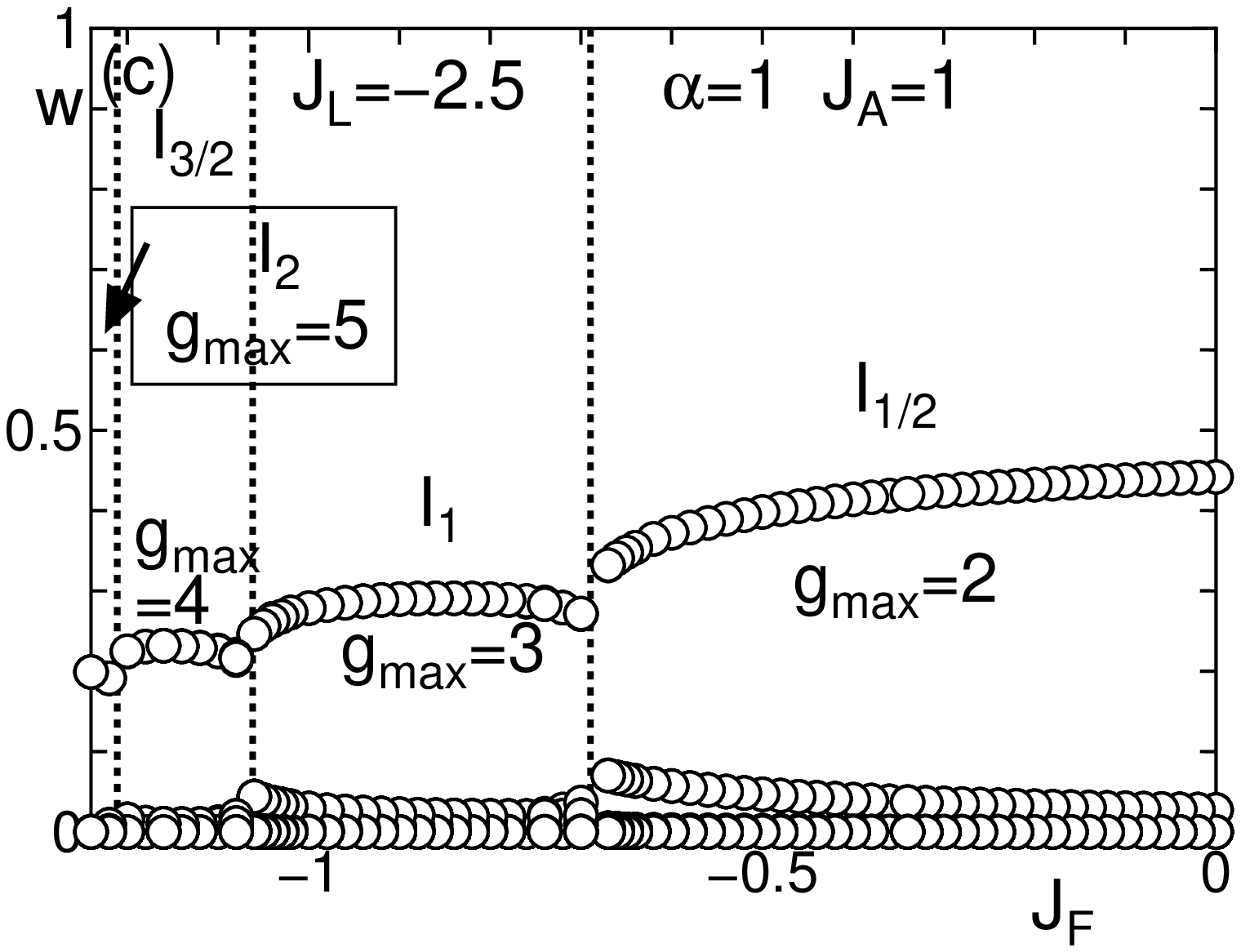}}
\caption{$\JF$ dependences of several large eigenvalues $w_{\alpha}$ of the density matrices $\rho_{\rm R(L)}$ and the degeneracy of the largest eigenvalue $g_{\rm max}$ with division A for  (a) $\Jl= -1.75$,  (b) $-2$, and (c) $-2.5$  with $\JA=1$ and $\alpha=1$. The vertical dotted lines are the phase boundaries determined in Ref. \citen{hts}. 
} 
\label{fig:enta}
\end{figure}
\begin{figure}[h!] 
\centerline{\includegraphics[width=7cm]{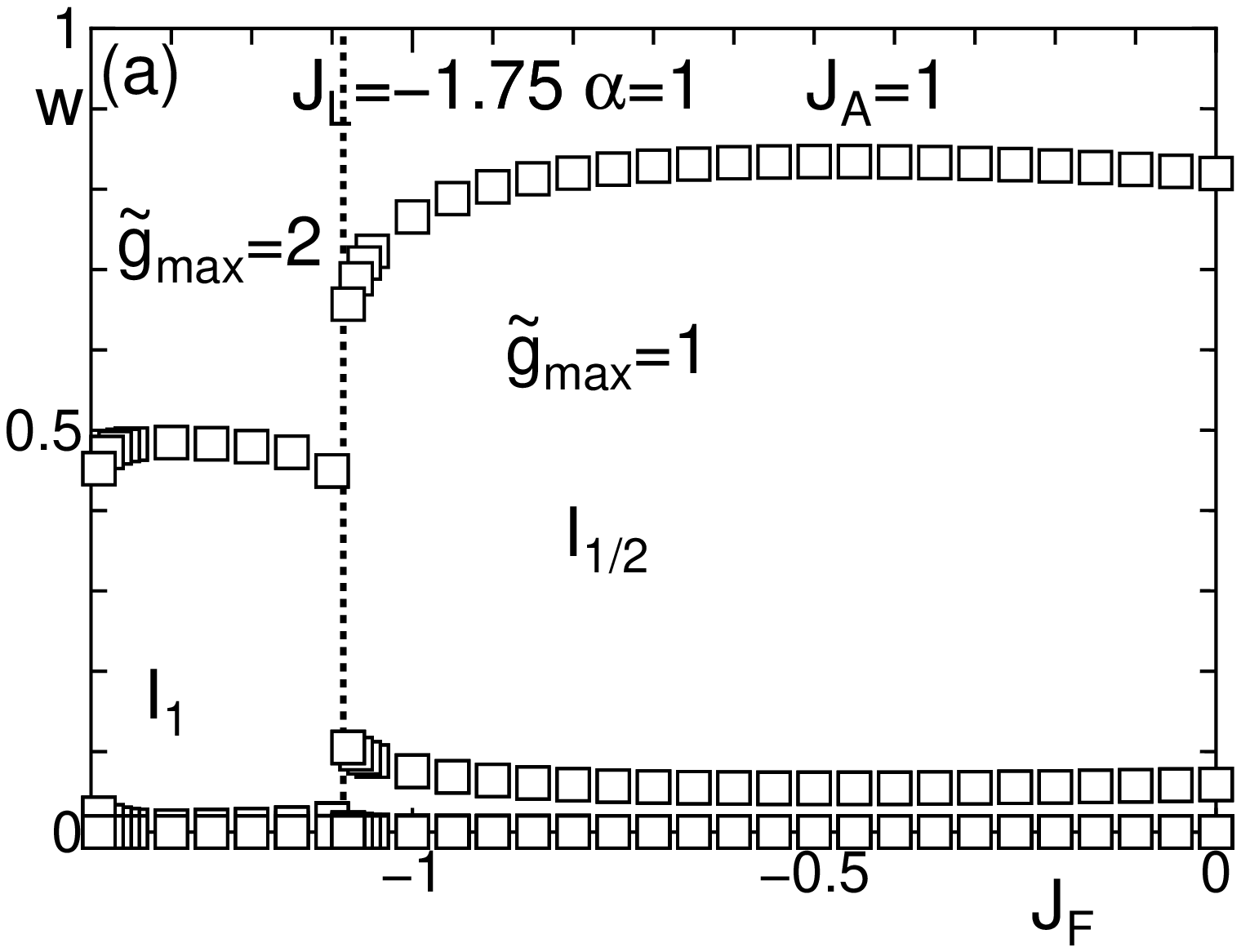}}
\centerline{\includegraphics[width=7cm]{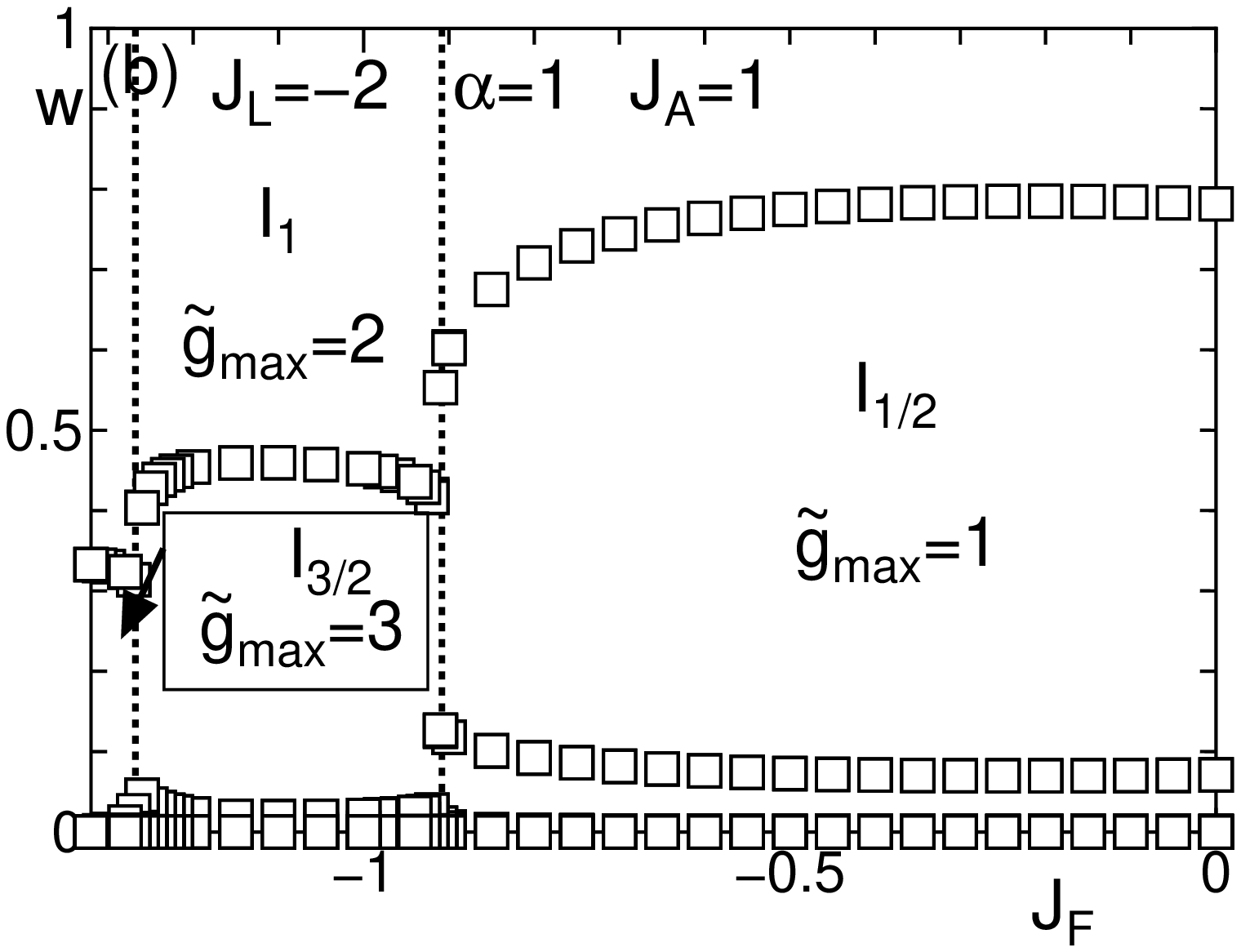}}
\centerline{\includegraphics[width=7cm]{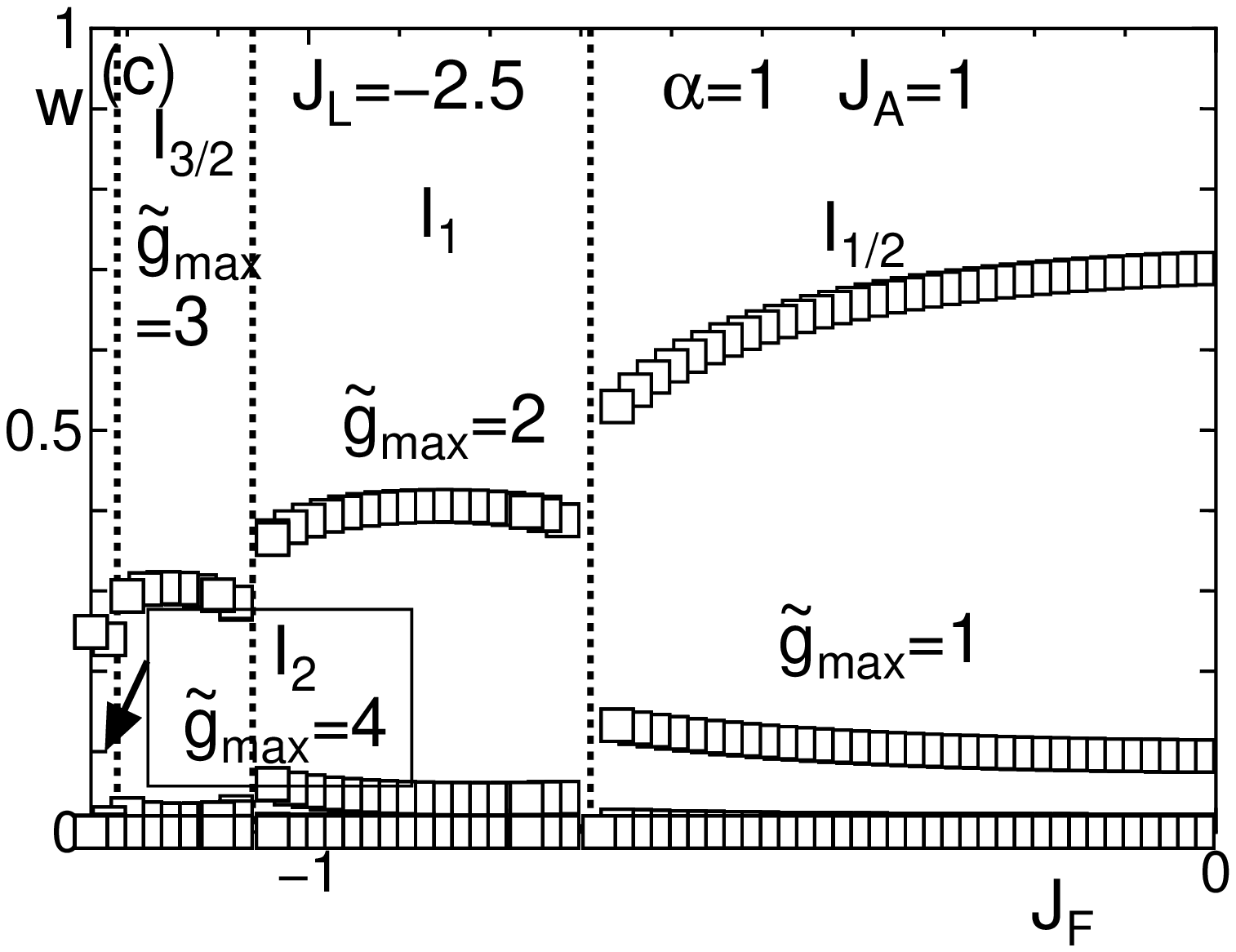}}
\caption{$\JF$ dependences of several large eigenvalues $w_{\alpha}$ of the density matrices $\rho_{\rm R(L)}$ and the degeneracy of the largest eigenvalue $\tilde{g}_{\rm max}$  with division B 
for  (a) $\Jl= -1.75$,  (b) $-2$, and (c) $-2.5$  with $\JA=1$ and $\alpha=1$. The vertical dotted lines are the phase boundaries determined in Ref. \citen{hts}. 
} 
\label{fig:entb}
\end{figure}
\begin{figure}[h!] 
\centerline{\includegraphics[width=7cm]{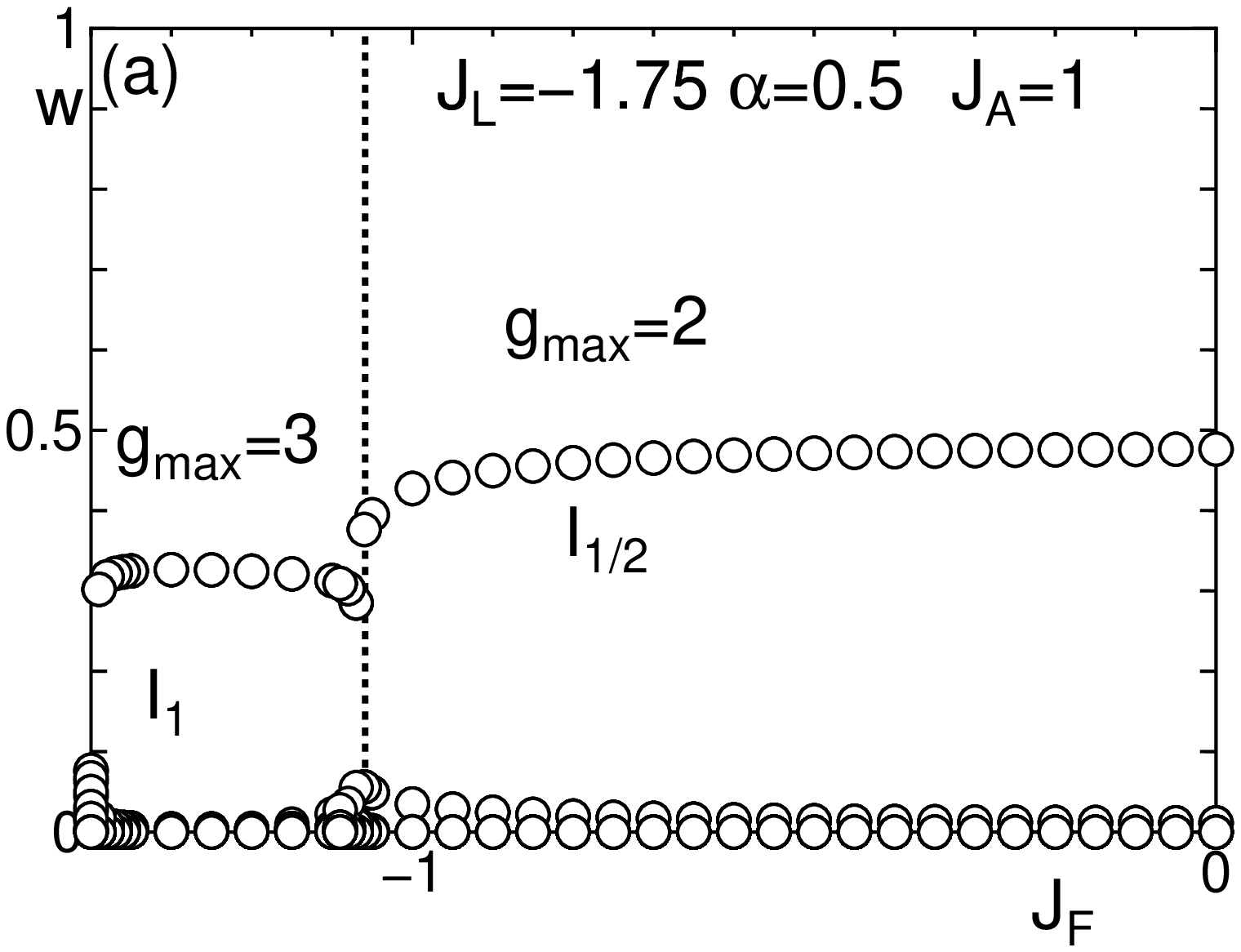}}
\centerline{\includegraphics[width=7cm]{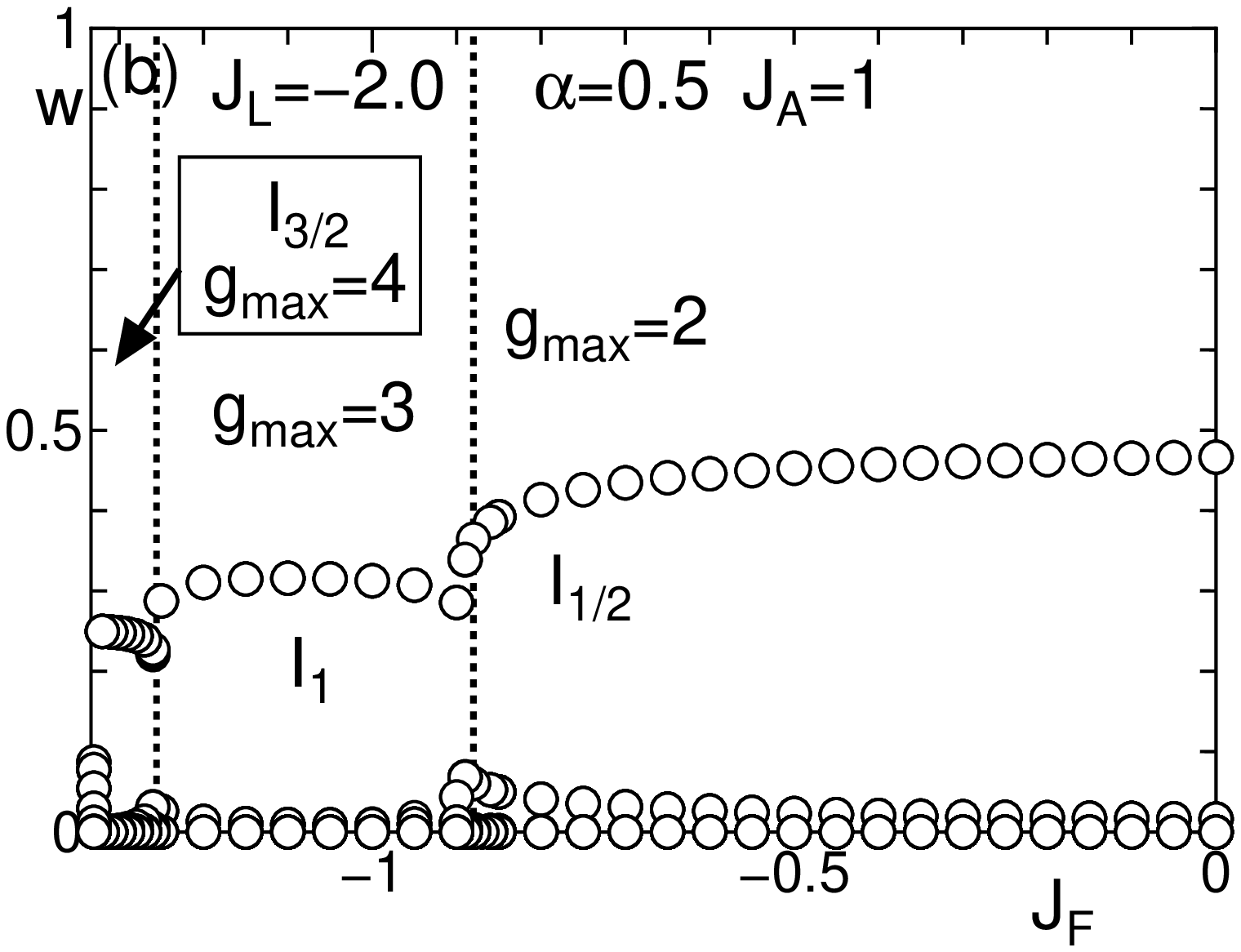}}
\centerline{\includegraphics[width=7cm]{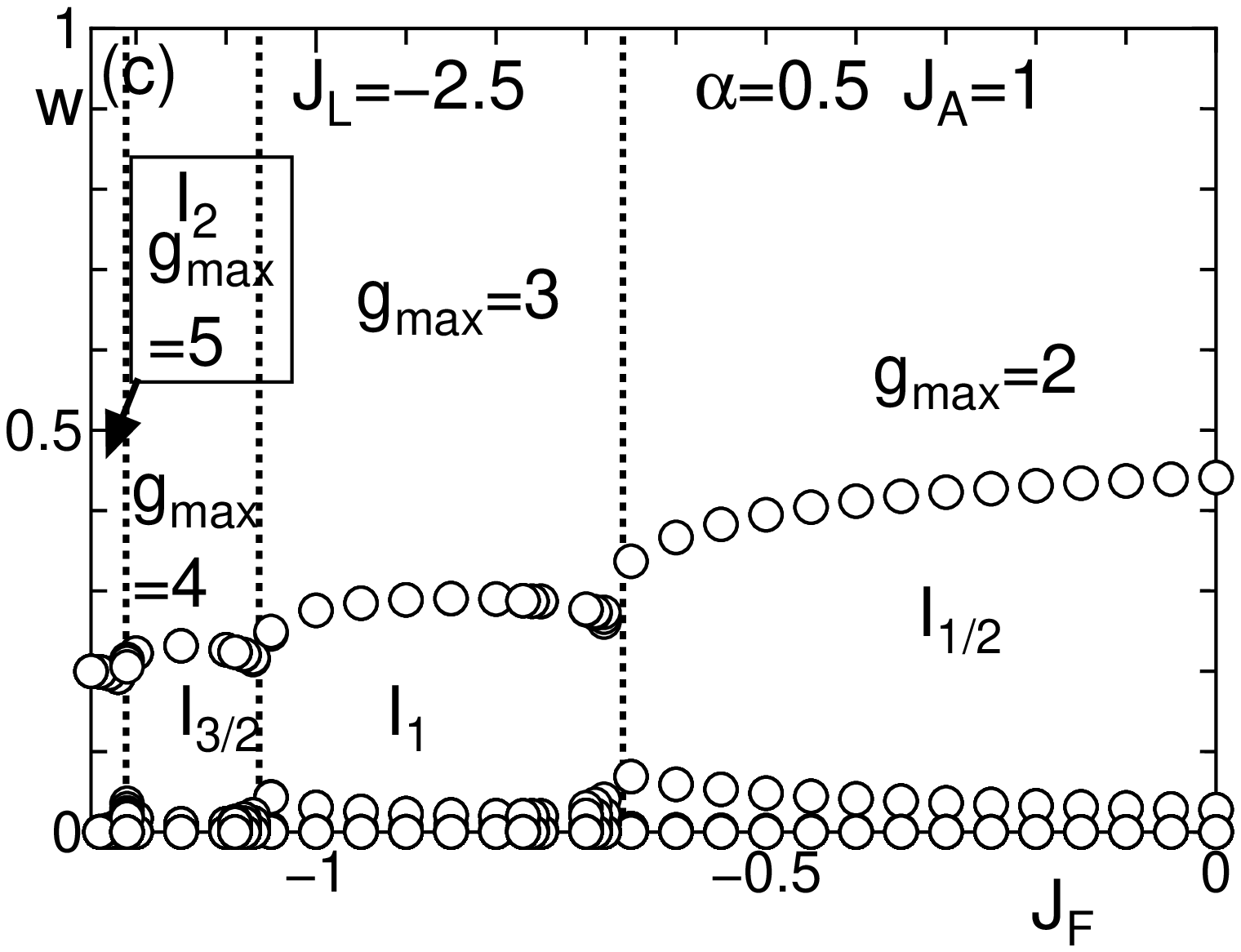}}
\caption{$\JF$ dependences of several large eigenvalues $w_{\alpha}$ of the density matrices $\rho_{\rm R(L)}$ and the degeneracy of the largest eigenvalue $g_{\rm max}$ with division A 
for  (a) $\Jl= -1.75$,  (b) $-2$, and (c) $-2.5$  with $\JA=1$ and $\alpha=0.5$. 
} 
\label{fig:ent05a}
\end{figure}
\begin{figure}[h!] 
\centerline{\includegraphics[width=7cm]{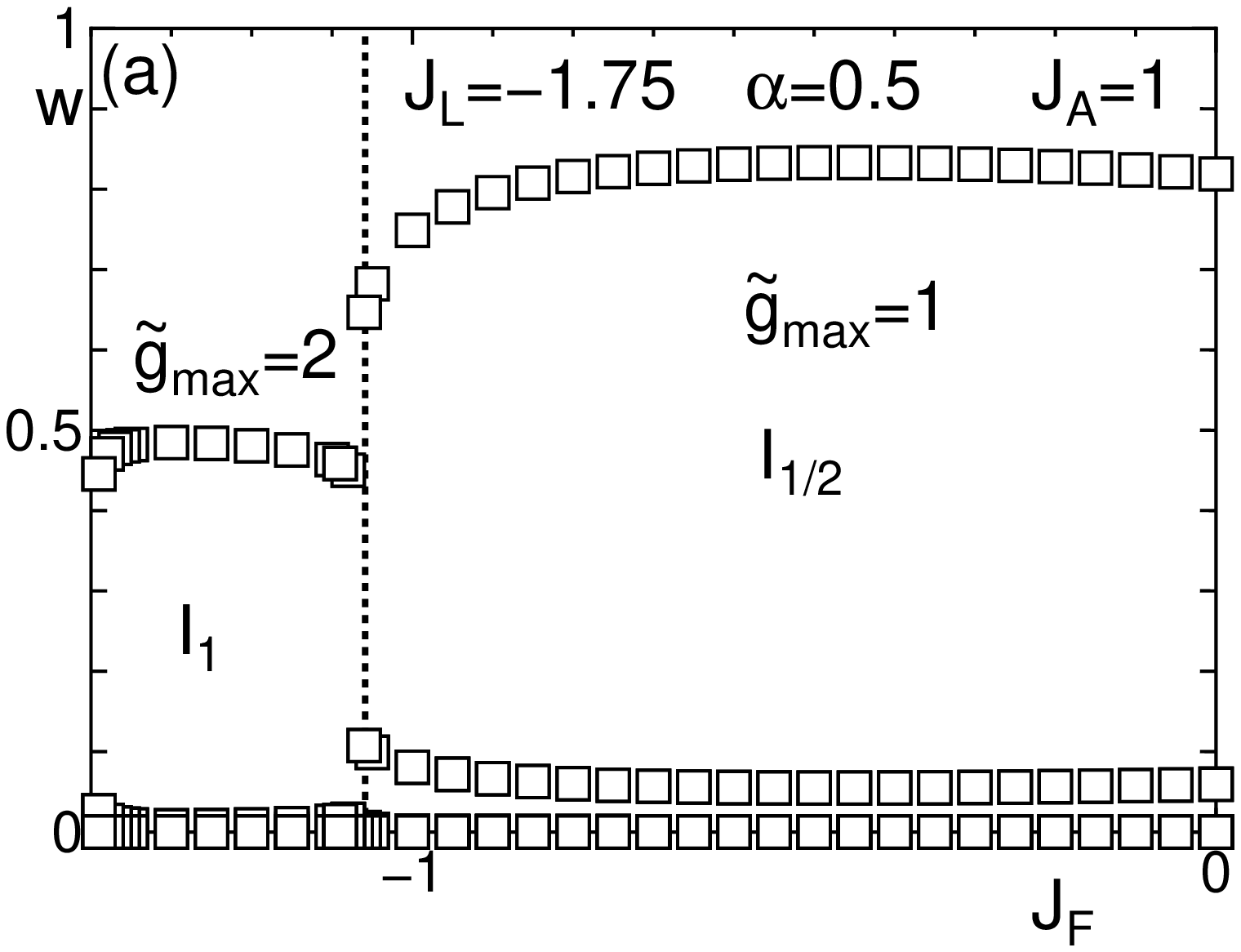}}
\centerline{\includegraphics[width=7cm]{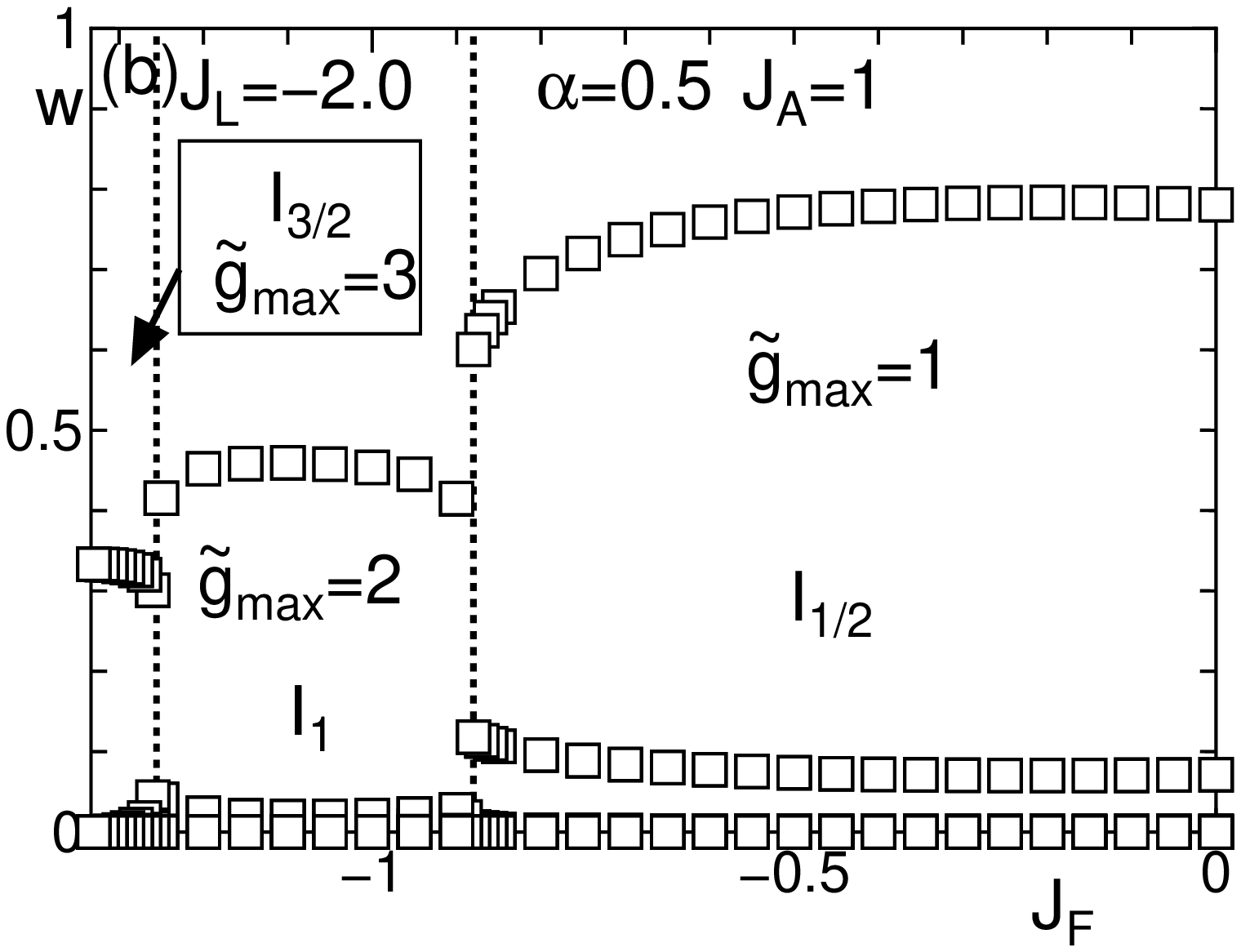}}
\centerline{\includegraphics[width=7cm]{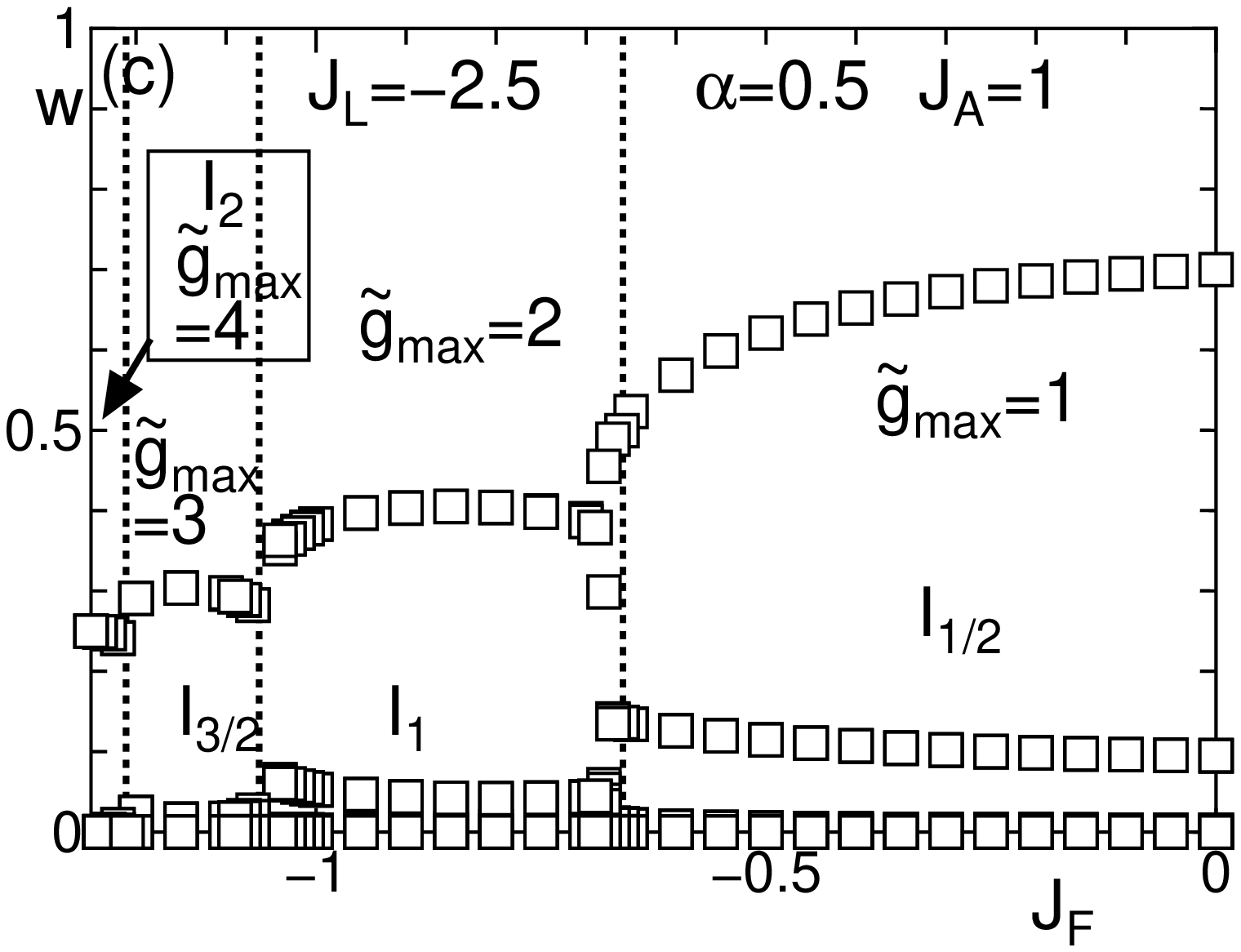}}
\caption{$\JF$ dependences of several large eigenvalues $w_{\alpha}$ of the density matrices $\rho_{\rm R(L)}$ and the degeneracy of the largest eigenvalue $\tilde{g}_{\rm max}$ with division B 
for (a) $\Jl= -1.75$,  (b) $-2$, and (c) $-2.5$ with $\JA=1$ and $\alpha=0.5$. 
} 
\label{fig:ent05b}
\end{figure}

\section{Entanglement Spectrum}

To calculate the ES, we divide the whole chain into left and right subsystems. 
The density matrices of the right and left subsystems, $\rho_{\rm R}$ and $\rho_{\rm L}$, are defined by
\begin{align}
\rho_{\rm R (L)} &= \Tr_{\rm L (R)}\ket{G}\bra{G},
\end{align}
where $\ket{G}$ is the ground state of the whole system and $\Tr_{\rm L(R)}$ implies the trace over the left (right) subsystem. The  eigenvalues $w_{\alpha}$  and eigenstates $\ket{\alpha}_{\rm R(L)}$ of the density matrix $\rho_{\rm R (L)}$ satisfy the eigenvalue equation
\begin{align}
\rho_{\rm R(L)}\ket{\alpha}_{\rm R(L)} &= w_{\alpha}\ket{\alpha}_{\rm R(L)}.
\end{align}
The set of eigenvalues $\{w_{\alpha}\}$ common to $\rho_{\rm R}$ and $\rho_{\rm L}$ forms an ES of the ground state of the whole system. Note that the ES depends on how the whole system is divided into subsystems.

We employ the iDMRG\cite{McCulloch2008,Schllwock2011,kh2016} method to calculate the ES.  We calculate the ES for both divisions A and B depicted in Fig. \ref{fig:cut}. 
 In Ref. \citen{kh2016}, it is pointed out that the degeneracies $g_{\rm max}$ for division A and $\tilde{g}_{\rm max}$ for division B of the largest eigenvalue of the density matrix are related to $\sed$ as $g_{\rm max}=2\sed+1$  for division A and $\tilde{g}_{\rm max}=2\sed$ for division B. We employ this relationship to identify the I$_{\sed}$ phase from the ES in general.

 In most cases, the number of  states kept in each subsystem is 240. In the neighborhood of the phase boundaries, it is increased  to 480 to ensure the convergence. For $\alpha=1$, several large eigenvalues $w_{\alpha}$ are shown in Fig. \ref{fig:enta} for division A and in Fig. \ref{fig:entb} for division B. The degeneracies of the largest eigenvalues are also shown in Fig. \ref{fig:enta} as $g_{\rm max}$ for division A and in Fig. \ref{fig:entb} as $\tilde{g}_{\rm max}$ for division B. 
We identify the I$_{\sed}$ phase by $g_{\rm max}=2\sed+1$ for division A. Then, for all I$_{\sed}$ phases, we find $\tilde{g}_{\rm max}=2\sed$ for division B. Compared with the edge spins at the open boundaries that can be affected by the local modulation of interactions, the degeneracy of the ES is an intrinsic bulk property. Physically, this degeneracy results from the virtual free  spins with the magnitude $\sed$ at the boundary between the left and right subsystems for division A. Those with the magnitude $\sed-1/2$ appear for division B.

The ES for $\alpha=0.5$ is shown in Fig. \ref{fig:ent05a} for division A and in Fig. \ref{fig:ent05b} for division B. The degeneracies of ES change at the dotted lines, which are the phase boundaries for $\alpha=0$. Thus, comparing these results with the results for $\alpha=0$ in Ref. \citen{kh2016} and for $\alpha=1$  in Figs. \ref{fig:enta} and \ref{fig:entb}, we may safely conclude that the phase boundaries and ES are both insensitive to $\alpha$.

The relationship between the degeneracies of ES ($g_{\rm max}$, $\tilde{g}_{\rm max}$) and valence bond structure is understood as follows: If $n_{\rm vb}$ valence bonds are cut by division A, $n_{\rm vb}$ virtual spins with $S=1/2$ emerge at the boundary. According to the valence bond solid structure proposed in Ref. \citen{hts}, these spins are located on the even-th (odd-th) sites on the left (right) half subsystem. Owing to the strong ferromagnetic coupling $\Jl$, these virtual spins are fully symmetrized with $\sed=n_{\rm vb}/2$ and $g_{\rm max}=2\sed+1$. For division B, the number of  virtual spins decreases by unity. Hence, we have $\tilde{g}_{\rm max}=2\sed$. Since the valence bonds are formed between virtual spins on even-th sites and those on odd-th sites in the bulk ground state, the symmetrization among the even-th spins results in the symmetrization among odd-th spins and vice versa. This explains why the ES are insensitive to  $\alpha$. 

\section{Summary and Discussion}
We employed the iDMRG method to calculate  the ES of the spin-1/2 ferromagnetic--antiferromagnetic alternating Heisenberg chain with alternating  next-nearest-neighbour interaction for divisions A and B.   We confirmed the presence of successive frustration-induced phase transitions between a series of SPT and trivial spin-gap phases for $\alpha=1$($\Delta$-chain) and $\alpha=0.5$ by the characterization of these phases in terms of  the ES. In addition, we found that the phase boundaries between these phases are insensitive to $\alpha$. The valence bond solid picture of each phase proposed in Ref. \citen{hts} is consistent with the present results irrespective of $\alpha$.

In the I$_{\sed}$ phase, the ES in division A behaves as if the whole chain consisted of spins with the magnitude $2\sed$. We may speculate on the physical origin of these effective spins in the following manner: For a large ferromagnetic $\Jl$, the spins connected by $\Jl$ are strongly correlated ferromagnetically with each other. In the nonmagnetic I$_{\sed}$ phase, however, this correlation does not extend over the whole chain owing to the frustration induced by  $\JA$ and $\JF$, which reduces the ferromagnetic correlation  between $\v{S}_i$ and $\v{S}_{i+2}$. In particular, the antiferromagnetic $\JA$ tends to break the long-range order dynamically rather than weaken the magnitude of the static ferromagnetic long-range order. Hence, the spins behave as an assembly of finite size clusters with a large but finite effective spin. Since this large effective spin is an extended object, the resulting valence bond solid state consists of the resonating state of valence bonds with various lengths as described in Fig. 14 of Ref. \citen{hts}. With further increase in $|\JF|$, the ferromagnetic correlation extends over the whole chain, resulting in the ferromagnetic ground state. Since the ES in the nonmagnetic phases are almost insensitive to $\alpha$, the structure of the wave function should  also be insensitive to $\alpha$. Hence, the picture described in Ref. \citen{hts} remains valid in the present case of $\alpha \neq 0$. Although this picture is consistent with the ES and the magnitudes of the edge spins, further investigation is desirable to confirm this picture more directly.

The investigations 
 in the present and preceding works\cite{hts,kh2016} are limited to the models with exact solutions on the ferromagnetic-nonmagnetic phase boundary.  In these works, the  numerical results suggesting the presence of successive transitions are  reinforced by the exact solutions.  Hence, to conclude that 
 the presence of the successive transitions is  one of the generic features of one-dimensional frustrated quantum ferromagnets, it is essentially important to show that  it persists even in the absence of exact solutions. Actually, similar successive transitions have been suggested in the frustrated mixed spin chains with side chains,\cite{htsdec} although the results were not conclusive owing to the limited system size. This model contains the \DCFFA \ as a limiting case. Another possible extension of the present model is the frustrated ladder with ferromagnetic legs and unequal antiferromagnetic diagonal bonds.\cite{hi,kato} This model reduces to the \FFAANNN, if one of the diagonal bonds vanishes. Studies of these models that have no exact solutions even on the ferromagnetic-nonmagnetic phase boundary are highly desirable in the future. 

Further theoretical and experimental searches for these exotic phases would be a promising subject in the field of frustrated quantum magnetism. Thus far, these transitions have not yet been observed experimentally. For experimental detection, it would be important to clarify how the difference in ground-state phase is reflected in the finite temperature behaviors of experimentally observable quantities such as magnetic susceptibility and specific heat. These are left for future studies.  

\acknowledgments
The author thanks S. C. Furuya for showing him the results of Ref. \citen{furuya} prior to publication and for fruitful discussion. Part of the numerical computation in this work has been carried out using the facilities of the Supercomputer Center, Institute for Solid State Physics, University of Tokyo, and   Yukawa Institute Computer Facility in Kyoto University. This work is  supported by JSPS KAKENHI Grant Number JP25400389.

\end{document}